\documentclass[linenumbers]{aastex631}
\shorttitle{Triple nucleus in NGC 6764}
\shortauthors{Heidt et al.}
\graphicspath{{./}{figures/}}

\begin{document}

\title{Detection of a triple nucleus in the Wolf-Rayet composite Sy 2 galaxy NGC 6764}

\author[0000-0002-0320-1292]{J. Heidt}
\affiliation{Landessternwarte, Zentrum f\"ur Astronomie der Universit\"at Heidelberg, K\"onigstuhl 12, 69117 Heidelberg, Germany}

\author[0000-0002-6716-4179]{F. Pozo Nu\~nez}
\affiliation{Astroinformatics, Heidelberg Institute for Theoretical Studies, Schloss-Wolfsbrunnenweg 35, 69118 Heidelberg, Germany}

\author[0009-0006-3864-7645]{N. Mackensen}
\affiliation{Landessternwarte, Zentrum f\"ur Astronomie der Universit\"at Heidelberg, K\"onigstuhl 12, 69117 Heidelberg, Germany}

\author[0009-0007-3538-9879]{M. Vukojevic}
\affiliation{Landessternwarte, Zentrum f\"ur Astronomie der Universit\"at Heidelberg, K\"onigstuhl 12, 69117 Heidelberg, Germany}

\author{D. Thompson}
\affiliation{LBT Observatory, University of Arizona, 933 N.~Cherry Ave, Tucson, USA}



\begin{abstract}
We report adaptive-optics $K_{s}$-band imaging of the composite Seyfert~2 and Wolf-Rayet galaxy NGC~6764 ($D=32$\,Mpc), obtained with LUCI/SOUL at the Large Binocular Telescope. 
The nucleus resolves into three compact sources with pairwise projected separations of 
$0.07^{\prime\prime}$--$0.2^{\prime\prime}$, corresponding to $11.0$--$30.9\,
\mathrm{pc}$.

All three have indistinguishable $H$, $K_{s}$, H$_{2}$, and Br$\gamma$ colours, so photometry alone cannot separate accretion from star formation. The system is consistent with a Seyfert nucleus and two star-forming regions, but a dual or triple AGN cannot be excluded.
In case of the latter, the separations would be two orders of magnitude smaller than in any confirmed or candidate AGN triplet reported to date. Diffraction-limited near-infrared integral-field spectroscopy is required to establish the nature of each component.
\end{abstract}




\section{Introduction} \label{sec:intro}

We report adaptive-optics imaging that resolves the core of NGC~6764 into three compact components with pairwise projected separations of only $11.0$--$30.9\,\mathrm{pc}$.
If all three host accreting black holes, this would be the most closely separated candidate AGN triplet known, by two orders of magnitude.

NGC 6764 ($z=0.0081$, $D\simeq32$\,Mpc) is a barred spiral that is simultaneously a Wolf-Rayet galaxy \citep{1982ApJ...261...64O} and a composite object in the BPT diagram \citep{1981PASP...93....5B}, hosting a Seyfert 2 nucleus together with two
starburst populations of $\sim5$ and $\sim30$\,Myr within the central few hundred parsecs \citep{2000ApJ...545..205S}.
\citet{2000ApJ...545..205S} report ROSAT HRI count-rate variations by a factor of $\sim2$ on a timescale of 7 days, direct evidence for a compact accreting source.
The question of which of the three resolved components is responsible remains to be established.

Systems with multiple active nuclei are rare, and those known are widely separated,
so on precisely this scale the composition of such a nucleus has remained untested.
Confirmed and candidate AGN triplets have mutual separations of kiloparsecs to tens of kiloparsecs \citep{2014Natur.511...57D,Pfeifle_2019,2026A&A...712A..80U}.
Parsec-scale black hole pairs have been identified in a handful of radio galaxies through VLBI (e.g. \citealt{Rodriguez_2006}), but no comparable triple system has been resolved.

\section{Observations and results}

\begin{figure*}

\includegraphics[width=1\textwidth]{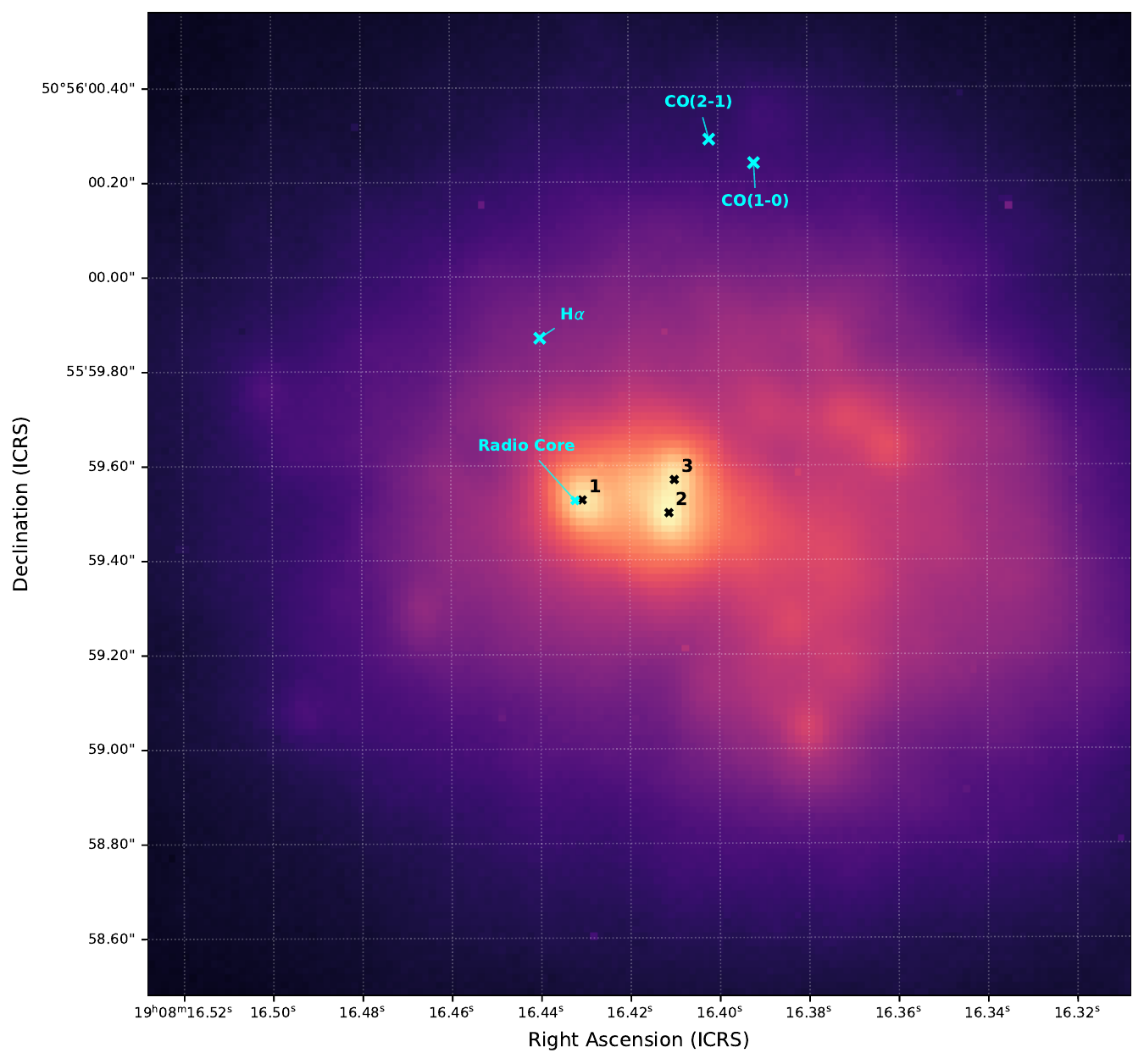}
\caption{Central $\sim2\farcs1\times2\farcs1$ region of NGC~6764 observed in the $K_{s}$ band with LUCI\,2/SOUL at the LBT. The astrometric solution was tied to the ICRS using three Gaia DR3 stars within the $30^{\prime\prime}$ LUCI\,2 field of view. Their catalogue positions, given at the reference epoch J2016.0, were propagated to the epoch of the observations, J2023, using the catalogued proper motions where available \citep{gaia2016,gaiadr3_summary}.
The nucleus resolves into three components (1--3) with pairwise separations of 
$0.19^{\prime\prime}$ (1--2), $0.20^{\prime\prime}$ (1--3), and $0.07^{\prime\prime}$ 
(2--3), corresponding to $28.8$, $30.9$, and $11.0\,\mathrm{pc}$ at $D=32\,\mathrm{Mpc}$, respectively.
The cyan crosses mark the VLBA core \citep{Kharb_2010} and the H$\alpha$ and CO peaks (\citealt{2007A&A...473..747L}). The astrometric accuracy is approximately $0.005^{\prime\prime}$ for the VLBA core, while the conservative absolute uncertainty of our Gaia-calibrated astrometry is $0.17^{\prime\prime}$. The uncertainties of the CO and H$\alpha$ positions are $0.2^{\prime\prime}$--$0.3^{\prime\prime}$.}
\label{fig1}
\end{figure*}

Diffraction-limited $H, K_s, H_2$ and $Br\gamma$ images of NGC 6764 were taken with the LUCI-instruments \citep{2003SPIE.4841..962S} supported by the AO-system SOUL 
\citep{2016SPIE.9909E..3VP}
at the Large Binocular Telescope (LBT) during the nights of June 30 / July 1 and July 5/6 2023, respectively.
The resulting $K_s$-band image of NGC~6764 is shown in Fig.~\ref{fig1}.
The core of NGC~6764 breaks into three distinct components labelled 1--3. Their pairwise separations are $0.19^{\prime\prime}$ (1--2), $0.20^{\prime\prime}$ (1--3), and $0.07^{\prime\prime}$ (2--3), corresponding to projected separations of $28.8$, $30.9$, 
and $11.0\,\mathrm{pc}$, respectively. These three nuclei are embedded in a fluffy environment and could represent the Seyfert 2 nucleus and the two starburst systems discussed by \cite{2000ApJ...545..205S}. 
The radio core detected by \cite{Kharb_2010} is positionally consistent with component~1 within the astrometric uncertainty of approximately $0.17^{\prime\prime}$. We therefore tentatively identify component~1 as the Seyfert~2 nucleus.
The colours of the three $K_s$ $\sim 15.9$ mag components derived via aperture photometry and GALFIT \citep{2010AJ....139.2097P} using the four filters are very similar and cannot be used to distinguish between AGN and starburst activity in them. 
Therefore, we cannot even rule out the possibility that we witness a dual AGN (one without detectable radio emission) and two co-spatial starburst systems or even a triple AGN.

\section{Outlook}
The three components have a maximum projected extent of approximately $31\,\mathrm{pc}$. For plausible enclosed masses of $10^{6}$--$10^{8}\,M_{\odot}$ and projected separations of approximately $11$--$31\,\mathrm{pc}$, the characteristic circular velocities are of order $10$--$200\,\mathrm{km \,s^{-1}}$. The corresponding dynamical timescales, $t_{\mathrm{dyn}}\sim(r^{3}/GM)^{1/2}$, range from approximately $0.05$ to $2.57\,\mathrm{Myr}$ and are therefore comparable to or shorter than the age of the youngest starburst \citep{2000ApJ...545..205S}.
This configuration is therefore likely to be dynamically short-lived.

Each of these interpretations carries distinct implications.
If the components are the Seyfert~2 nucleus and two super star clusters, NGC 6764 becomes a system in which AGN and starburst feedback can be traced spatially on scales of approximately $11$--$31\,\mathrm{pc}$, where the two processes compete directly for the same gas.
If two or three components host accreting black holes, this would be the smallest 
separation system be detected by direct imaging.

Discriminating between these scenarios requires spectroscopy with an IFU at the spatial resolution of the imaging. 
Near-infrared coronal lines ([Si\,\textsc{vi}]\,1.963\,$\mu$m, [Ca\,\textsc{viii}]\,2.32\,$\mu$m) cannot be produced by stellar photoionisation and would identify an AGN in each component independently of the degenerate continuum colours; broad He\,\textsc{i}\,1.083\,$\mu$m and Pa$\beta$ would yield
single-epoch virial masses and Eddington ratios; the Br$\gamma$/H$_{2}$/[Fe\,\textsc{ii}] ratios separate photoionisation from shocks and outflows, and relative radial velocities accurate to $\sim10$\,km\,s$^{-1}$ test directly whether the components are bound. 
Whatever it is, these observations allow to study the sphere of influence 
(about 10-30 pc for a \(10^{7}\) -- \(10^{8}\,M_\odot\) 
supermassive black hole) in Seyfert 2 galaxies.
At $\delta\simeq+50\degr$ this is achievable only with OSIRIS at Keck.

\begin{acknowledgments}
The LBT is an international collaboration among institutions in the United States and Europe. At the time data were acquired for this research, LBT Corporation Members were the University of Arizona on behalf of the Arizona Board of Regents; Istituto Nazionale di Astrofisica, Italy; LBT Beteiligungsgesellschaft, Germany, representing the Max-Planck Society, the Leibniz Institute for Astrophysics Potsdam, and Heidelberg University; and The Ohio State University, representing The Ohio State University, University of Notre Dame, University of Minnesota, and University of Virginia.  This research used the facilities of the Italian Center for Astronomical Archives (IA2) operated by INAF at the Astronomical Observatory of Trieste. Observations have benefited from the use of ALTA Center (alta.arcetri.inaf.it) forecasts performed with the Astro-Meso-Nh model. Initialization data of the ALTA automatic forecast system come from the General Circulation Model (HRES) of the European Centre for Medium Range Weather Forecasts.
This work was supported in part by the German federal department for education and research (BMBF) 
under the project numbers 05 AL2VO1/8, 05 AL2EIB/4, 05 AL2EEA/1, 05 AL2PCA/5, 05 AL5VH1/5, 05 AL5PC1/1 and 05 A08VH1. 
FPN gratefully acknowledges the generous and invaluable support of the Klaus Tschira Foundation.
FPN acknowledges funding from the European Research Council (ERC) under the European Union's Horizon 2020 research and innovation program (grant agreement No 951549).
\end{acknowledgments}

%








\bibliography{NGC6764_RNAAS}{}
\bibliographystyle{aasjournal}



\end{document}